\documentclass[%
 rmp,a4paper,twocolumn,% jmp
 amsmath,amssymb,superscriptaddress
reprint
]{revtex4-2}
\usepackage{graphicx}% Include figure files
\usepackage{dcolumn}% Align table columns on decimal point
\usepackage{bm}% bold math
\usepackage[T1]{fontenc}
\usepackage{mathptmx,xcolor}
\DeclareMathAlphabet\mathbfcal{OMS}{cmsy}{b}{n}

\renewcommand{\[}{\begin{equation}}
\renewcommand{\]}{\end{equation}}
\def\bea{\begin{eqnarray}}
\def\eea{\end{eqnarray}}
\def\nn{\nonumber\\}

\newcommand{\emi}[1]{{\rm e}^{-i #1}}
\newcommand{\ei}[1]{{\rm e}^{i #1}}

\newcommand{\PP}{{\hat P}}

\newcommand{\A}{{\bf A}}
\renewcommand{\j}{{\bf j}}
\newcommand{\p}{{\bf p}}

\renewcommand{\k}{{\bf k}}

\newcommand{\kk}{\boldsymbol\kappa}
\newcommand{\pp}{\boldsymbol\pi}
\newcommand{\EE}{\protect{\cal E}}
\def\EEE{\protect{\mathbfcal E}}
\renewcommand{\v}{{\bf v}}
\renewcommand{\r}{{\bf r}}
\newcommand{\R}{{\bf R}}
\renewcommand{\P}{{\bf P}}

\newcommand{\dda}{\partial_{\kappa}}
\newcommand{\ddb}{\partial_{\lambda}}

\newcommand{\equ}[1]{Eq.~(\ref{#1})}
\newcommand{\eqs}[2]{Eqs.~(\ref{#1}) and (\ref{#2})}

\def\bra#1{\langle#1\vert}
\def\ket#1{\vert#1\rangle}
\def\ev#1{\langle#1\rangle}
\def\me#1#2#3{\langle#1| \, #2 \, |#3\rangle}
\def\runtime{(\the\time)\qquad\the\month/\the\day/\the\year}% get current time
\def\today
 {\count10=\year\advance\count10 by -2000 \number\day--\ifcase
  \month \or Jan\or Feb\or Mar\or Apr\or May\or Jun\or
             Jul\or Aug\or Sep\or Oct\or Nov\or Dec\fi--\number\count10}

\def\hour{\count10=\time\count11=\count10
\divide\count10 by 60 \count12=\count10
\multiply\count12 by 60 \advance\count11 by -\count12\count12=0
\number\count10 :\ifnum\count11 < 10 \number\count12\fi\number\count11}
\begin{document}

\title{{\bf Adiabatic observables and Berry curvatures in insulators and metals}}
\author{Raffaele Resta}

%\email{resta[at]iom.cnr.it}
\affiliation{CNR-IOM -- Istituto Officina dei Materiali, National Research Council of Italy, 34149 Trieste, Italy}
\affiliation{Donostia International Physics Center, 20018 San Sebasti{\'a}n, Spain}
%\date{\ttoday}
\date{\sf DRAFT: run through \LaTeX\ on \today\ at \hour}

\begin{abstract} A sharp definition of what ``adiabatic'' means is given; it is then shown that the time-dependent expectation value of a quantum-mechanical observable in the adiabatic limit can be expressed---in many cases---by means of the appropriate Berry curvature. Condensed-matter observables belonging to this class include: induced macroscopic polarization, Born effective charges in insulators and in metals, quantized Faraday charges in electrolytes, and linear dc conductivities (longitudinal and transverse). Spontaneous polarization is instead expressed by means of a related geometrical entity: the Berry phase.
Remarkably, the adiabatic limit is well defined even in metals, despite the absence of a spectral gap therein. For all of the above observables the explicit Berry-curvature expressions are derived in a general many-body setting, which also allows for compact and very transparent notations and formulas. Their conversion into band-structure formulas in the independent-electron crystalline case is straightforward.
\end{abstract}

\maketitle
%\nopagebreak

%\begin{widetext}
\tableofcontents

\section{Introduction}

The term ``adiabatic'' intuitively refers to the time evolution of a system which is driven out of equilibrium by an ``infinitely slow'' perturbation. Consider a quantum system in its stationary ground state, whose Hamiltonian acquires a time dependence: I am addressing here expressions for some given observables which are exact in the adiabatic limit. Suppose the Hamiltonian is driven by a parameter $\lambda(t)$: the adiabatic limit is then defined as the limit $\dot\lambda(t) \rightarrow 0$; the adiabatic evolution of a given observable coincides with the exact one to linear order in $\dot\lambda(t)$ and neglects terms of order $\ddot\lambda(t)$ and higher. Equivalently, when the acquired time dependence of the Hamiltonian is harmonic at frequency $\omega$, the adiabatic evolution of a given observable is exact to linear order in $\omega$ and differs from the exact evolution by  terms of order $\omega^2$ and higher. The expression "infinitely slow" has therefore the actual meaning of "for infinitesimal $\dot\lambda(t)$".

The present Review derives explicit expressions for the adiabatic evolution of several observables; all of them only require the solution of time-independent Schr\"odinger equations, while going beyond the adiabatic limit (i.e. including terms of order $\ddot\lambda(t)$  and beyond) would require addressing the genuine time-dependent Schr\"odinger equation and devising suitable approximations for its solution. In some (though not all) of the observables discussed here the time evolution of the Hamiltonian owes to nuclear displacements. Within the Born-Oppenheimer approximation  (also called Born-Huang)  the equations of motion of a system of electrons and nuclei are decoupled and one diagonalizes the electronic Hamiltonian at fixed values of the nuclear coordinates, collectively identified with the parameter $\lambda$. The time evolution of the electronic wavefunctions and of the corresponding observables is then approximated with their adiabatic evolution: the latter only is addressed here. The Born-Oppenheimer approximation becomes exact for infinitely slow ionic motions, while at finite frequencies it is---as the name says---an approximation;
 whether it can be applied as such to a given molecule, liquid, or solid and to a given phenomenon depends of course on its accuracy in the case study.

Adiabaticity is deeply related to electron transport theory. This is clearly demonstrated by a well known apparent paradox: continuity equation is obviously violated if both the charge density and the current density are evaluated as the expectation values of the corresponding operators on the instantaneous adiabatic ground state. In the simple case of a time-reversal (T) invariant system the current so evaluated vanishes at all times, while the charge density is clearly time-dependent. The missing term is provided by adiabatic-transport theory, which owes to the work of  \textcite{Thouless83} and \textcite{Berry84}. The focus of the present Review is on interacting many-electron systems; in this case the theory owes to a classical paper of \textcite{Niu84}. Adiabatic transport theory is rooted in quantum geometry by means of quantities which today are commonly called Berry curvatures \cite{rap_a20,Xiao10,Vanderbilt}.

This Review is devoted to addressing geometry-wise a number of condensed-matter observables: induced macroscopic polarization, Born effective charges in solids and liquids (insulating and metallic), Faraday's transported charges in electrolytes, and linear dc conductivities (both longitudinal and transverse). 
All these observables are adiabatic (according to the above definition), and all will be cast in terms of the appropriate Berry curvature; additionally,  spontaneous polarization is also addressed. It is an adiabatic observable formulated in terms of a different geometrical entity: the Berry phase.

The geometrical nature of the above observables is nowadays obvious in some cases (macroscopic polarization, Born charges in insulators, transverse dc conductivity), little known in one case (Faraday charges), and probably unexpected in the metallic cases (longitudinal dc conductivity and Born charges).
Addressing adiabatic responses in metals may look surprising, given the absence of a spectral gap therein (in the thermodynamic limit); nonetheless I am going to show that it is legitimate within Kohn's conceptual approach to dc conductivity \cite{Kohn64,Scalapino93,rap157} adopted here. In some special cases a geometrical observable may become topological: its value is then expressed by a universal constant times an integer number. In such cases  the observable is in principle measurable with infinite precision, the limit being actually set by the known accuracy of the universal constant. As for the present Review, I only deal with two topological observables: Faraday  charges in electrolytes (Sec. \ref{sec:Faraday}), and Hall conductivity in a two-dimensional insulator (Sec. \ref{sec:ahc}).

A  large part of the literature on related topics is formulated at the independent-electron level, in a band structure framework: this is the case of macroscopic polarization \cite{Vanderbilt}, Born charges---in insulators \cite{Baroni01} and in metals \cite{Dreyer22}---and of transverse dc conductivity \cite{Nagaosa10}. Here instead I address an interacting many-electron system, possibly noncrystalline; the present results generalize some of the established independent-electron ones, but also allow for very compact notations and formulas and very transparent derivations. A simple prescription is given for converting our compact formulas into their band-structure counterparts, i.e. into Fermi-volume integrals.

This Review is organized as follows. Sec. \ref{sec:funda}  addresses the most general quantum system: in Sec \ref{sec:berry} the Berry connection and Berry curvature are defined and their main properties are outlined; in Sec. \ref{sec:adia} it is shown  how the appropriate Berry curvature yields the time-evolution of a generic observable in the adiabatic limit.
From Sec. \ref{sec:transport} onwards this Review addresses a many-electron system.
Sec. \ref{sec:H}  sets the notations for the Hamiltonian and the eigenstates of an interacting $N$-electron system. In Sec. \ref{sec:curr} it is proved that the charge density and the current density---both evaluated in the adiabatic limit---conserve continuity equation; this applies on the same ground  to molecules and to condensed matter. From this point onwards this Review specializes to solids and liquids, whose many-body boundary conditions are specified in Sec. \ref{sec:mol}. Sec.  \ref{sec:mcurr} addresses \textcite{Kohn64} approach to macroscopic charge current and its adiabatic evolution, in the case of both metals and insulators. Sec. \ref{sec:pol} is devoted to the theory of polarization (originally defined ``modern'') in its many-body formulation. Following the historical development, I first address induced polarization (Sec. \ref{sec:ind}), then polarization differences (Sec. \ref{sec:diff}), and ultimately polarization ``itself'' (Sec. \ref{sec:itself}). The multivalued nature of the latter is further clarified in Sec. \ref{sec:multi}. The main concepts of adiabatic charge transport are then applied  to liquid electrolytes in Sec. \ref{sec:Faraday}; therein it is shown that Faraday laws of electrolysis owe to the topological quantization of charge transport. Sec  \ref{sec:Drude} provides a very transparent derivation of the famous Kohn's formula for longitudinal dc conductivity, showing that even this property can be recast in terms of a Berry curvature.
The many-body Berry curvature provides a very compact expression for the Born effective charges, which applies to both insulators and metals: this is shown in  Sec. \ref{sec:Drudem}; in the metallic case the Born charges are related to longitudinal conductivity by a sum rule recently found---in an independent-electron framework---by \textcite{Dreyer22}. Sec. \ref{sec:ahc} shows that the present general setting allows for a one-line proof of the expression for the anomalous Hall conductivity. Its topological quantization---in the special case of a two-dimensional insulator---is also proved. Some conclusions are drawn in Sec. \ref{sec:conclu}. 

Appendix \ref{sec:cont} proves continuity equation for an interacting $N$-electron system in the exact case when both potentials (scalar and vector) depend on time. The following Appendices convert the many-body formulas into their independent-electron counterparts: in B for a generic case, in C for a condensed matter system (both crystalline and noncrystalline).

\section{Quantum geometry and adiabaticity}  \label{sec:funda}

\subsection{Berry connection and Berry curvature} \label{sec:berry}

Let $\hat{H}$ be a time-independent Hamiltonian and let $\ket{\Psi_n}$ be its eigenstates with eigenvalues $E_n$; when  $\hat{H}$ depends on a generic real parameter $\lambda$,  the eigenstates and eigenvalues are parameter-dependent as well. The ground state is assumed to be nondegenerate for all $\lambda$; its Berry connection is then defined as \[ {\cal A}_\lambda = i \ev{\Psi_0|\partial_\lambda \Psi_0} , \]
and measures the infinitesimal phase variation of the eigenstate when $\lambda$ is varied: \[ d\phi = - \mbox{Im ln} \ev{\Psi_{0\lambda} | \Psi_{0\lambda+d\lambda} } = {\cal A}_\lambda d \lambda.  \label{berry} \] Since the eigenstate $\ket{\Psi_{0\lambda}}$ is arbitrary by a $\lambda$-dependent phase factor, the connection is gauge-dependent and cannot be endowed---as such---with any physical meaning; because of this its role has been overlooked for many years.

In a milestone paper \textcite{Berry84} realized that, whenever the Hamiltonian is periodical in $\lambda$, the loop integral of the connection \[ \gamma = \oint d \phi \] is a gauge-invariant phase---called nowadays Berry phase---defined modulo $2\pi$. The basic tenet of Berry's paper is that any gauge-invariant quantum-mechanical expression corresponds in principle to an observable. Since then, several observables having the nature of a Berry phase have been identified in electronic structure theory \cite{rap_a20,Xiao10,Vanderbilt} and in many other domains. 

Next let us suppose that $\hat{H}$ also depends on a second real parameter $\kappa$;
the ground state is assumed to be nondegenerate for all $(\kappa,\lambda)$. 
The Berry curvature is by definition
\[ \Omega(\kappa,\lambda) =  \partial_\kappa {\cal A}_\lambda - \partial_\lambda {\cal A}_\kappa   = - 2 \,\mbox{Im } \ev{\dda \Psi_0 | \ddb \Psi_0} , \label{curva}\] 
%\bea \Omega(\kappa,\lambda) &=& i ( \ev{\dda \Psi_0 | \ddb \Psi_0} - \ev{\ddb \Psi_0 | \dda \Psi_0} ) \nn  &=& - 2 \,\mbox{Im } \ev{\dda \Psi_0 | \ddb \Psi_0} ; \label{curva}\eea 
and is gauge-invariant. When the parameters are time-dependent, the curvature
is a quasi-static quantity, in the sense that its definition requires solely the instantaneous ground eigenstate; yet it encodes the lowest-order effect of the excited states on the adiabatic evolution. 

The parameters may have various physical interpretations and different dimensions; the curvature has the inverse dimensions of the product $\kappa \lambda$. For a macroscopic homogeneous system $\Omega(\kappa,\lambda)$ is an extensive quantity.

The Berry curvature admits a sum-over-states (i.e. Kubo) formula, first displayed in the original Berry's paper:
\[ \Omega(\kappa,\lambda) = -2 \, \mbox{Im}  \sum_{n \neq 0} \frac{\ev{\Psi_0 | \dda\hat{H} | \Psi_n }\ev{\Psi_n | \ddb\hat{H} | \Psi_0 }}{(E_0 - E_n)^2} . \label{sum} \]  The formula perspicuously shows how the Berry curvature encodes---to lowest order---the effect of the excited states on the ground state when the quantum system is transported in the parameter space. \equ{sum}
has also the virtue of showing that the curvature becomes ill defined whenever the ground state is degenerate with the first excited state. 

The curvature is a geometrical quantity defined by the evolution of the ground-state projector \[ \PP = \ket{\Psi_0}\bra{\Psi_0} \label{proj}\] in the parameter space, and can equivalently be expressed directly in terms of $\PP$:  \[ \Omega(\kappa,\lambda) = i \, \mbox{Tr } \{ \PP \,[ \,\dda \PP, \ddb \PP\,] \} , \label{prcurv} \] where ``Tr'' indicates the trace over the Hilbert space. The form of \equ{prcurv} explicitly displays the gauge-invariance of $\Omega$:
a gauge-transformation modifies $\ket{\Psi_0}$ by a phase factor, ergo leaves $\PP$ and the curvature unchanged. 

\subsection{The adiabatic evolution of an observable} \label{sec:adia}

Let us consider the ground-state expectation value $O$ of a generic quantum-mechanical observable $\hat{O}$:  \[ O \equiv  \me{\Psi_0}{\hat{O}}{\Psi_0} . \] Suppose that the Hamiltonian acquires a time dependence by means of a parameter $\lambda(t)$: we address the time evolution of $O$.
The adiabatic theorem, in its textbook formulation \cite{Sakurai}, states that if a system is in a given nondegenerate eigenstate at time $t=0$, it remains in the instantaneous eigenstate if the perturbation is infinitely slow, and  no level is crossed in the time evolution.

The first step is therefore  adopting \[ O(t) \simeq  \me{\Psi_0}{\hat{O}}{\Psi_0} , \label{naif} \] where now $\ket{\Psi_0}$ is evaluated at $\lambda=\lambda(t)$,  thus acquiring  a parametric dependence on time. Depending on the nature of the observable \equ{naif} may be incomplete:  for instance if the quantum system is T-invariant and if $\hat{O}$ is imaginary in the Schr\"odinger representation (as e.g. a current or an angular momentum), then \equ{naif} yields identically zero, while there is no physical reason for $O(t)$ to vanish.

In the 1980s it was realized---both in condensed matter physics \cite{Thouless83,Niu84} and in quantum chemistry \cite{Nafie83,Stephens85}---that an extra term, proportional to $\dot\lambda(t)$, has to be added to \equ{naif} in order to make it exact in the adiabatic limit  for any choice of the operator $\hat{O}$. In the following I give an abridged derivation of the \textcite{Niu84} result, for the special case where $\hat{O}$ can be written as the derivative of the Hamiltonian with respect to some parameter $\kappa$: \[ \hat{O} = \partial_{\kappa} \hat{H} . \label{rap} \] 

Let us  the consider first the exact time evolution of $O(t)$. Let $\hat{H_t}$ be  a time-dependent Hamiltonian, and $\ket{\Psi}$ one of its solutions: \[ \hat{H}_t \ket{\Psi} = i \hbar \ket{\dot\Psi} . \] The time-dependent energy is \[ E(t) = \me{\Psi}{\hat{H}_t}{\Psi} ,\label{energy} \] and its $\kappa$-derivative is  \[ \partial_{\kappa} E(t) =  \me{\Psi}{\hat{O}}{\Psi} + i\hbar \, (\, \ev{\partial_{\kappa}\Psi|\dot\Psi} - \ev{\dot\Psi|\partial_{\kappa}\Psi} \,) . \label{exact} \] The {\it exact} time evolution of the observable's expectation value is therefore:  
\[ O(t) \equiv \me{\Psi}{\hat{O}}{\Psi}  = \partial_{\kappa} E(t)  + 2 \, \hbar \,\mbox{Im } \ev{\partial_{\kappa}\Psi|\dot\Psi}  . \label{exact2} \] 
%\bea O(t) &\equiv& \me{\Psi}{\hat{O}}{\Psi} \nn &=&\partial_{\kappa} E(t) - i\hbar \, (\, \ev{\partial_{\kappa}\Psi|\dot\Psi} - \ev{\dot\Psi_\kappa | \partial_{\kappa}\Psi} \,) . \label{exact2} \eea 

Next  we suppose as above
that the time dependence of $\hat{H_t}$ owes to  a slow parameter $\lambda$ entering it. As we previously did for \equ{naif}, we replace the exact quantities (energy and state vector) in \equ{exact2} with the instantaneous ones evaluated at $\lambda=\lambda(t)$. We thus 
get the adiabatic evolution of $O(t)$ in terms of the Berry curvature as \[ O(t) = \partial_{\kappa} E_0 - \hbar \, \Omega(\kappa,\lambda)\, \dot\lambda(t) , \label{NT} \] where $\ket{\Psi_0}$, $E_0$, and the curvature depend implicitly on time. 
To the best of the author's knowledge, a similar formulation first appeared in the \textcite{Niu84} paper; the original derivation also  explicitly shows that the exact time evolution of $O(t)$ differs from \equ{NT} by terms proportional to $\ddot\lambda(t)$ and higher.
For a time-independent $\lambda$ \equ{NT}  coincides with the Hellmann-Feynman theorem: therefore the equation can be regarded as the adiabatic generalization of it.

One may also rewrite \equ{NT} as \[ O(t) =  \me{\Psi_0}{\hat{O}}{\Psi_0} - \hbar \, \Omega(\kappa,\lambda)\, \dot\lambda(t) ,  \label{NTT} \] 
and consider the more general case where the time-independent operator $\hat O$ {\it cannot} be expressed as an Hamiltonian derivative. Then the adiabatic evolution of ${O(t)}$ obtains by replacing the curvature in \equ{NTT} with the corresponding Kubo formula, \equ{sum}: \bea {O(t)} &=& \me{\Psi_0}{\hat{O}}{\Psi_0} \nn &+& 2 \hbar \dot\lambda(t) \, \mbox{Im}  \sum_{n \neq 0} \frac{\ev{\Psi_0 | \hat{O} | \Psi_n }\ev{\Psi_n | \ddb\hat{H} | \Psi_0 }}{(E_0 - E_n)^2} , \label{sum2} \eea where once more the eigenstates and the eigenvalues are evaluated at $\lambda(t)$.

Finally it is worth observing that, depending on the nature of the operator $\hat{O}$, only one of the two terms in \eqs{NTT}{sum2} may be nonvanishing. A perspicuous example is in Sec. \ref{sec:mcurr}: the first term accounts for the adiabatic evolution of the density, while the second term accounts for the adiabatic evolution of the current.

\section{Adiabatic charge transport} \label{sec:transport}

\subsection{The Hamiltonian} \label{sec:H}

In this work the above very general quantum-mechanical results will be  adopted to deal with an interacting $N$-electron system, whose  time-independent Schr\"odinger equation is \bea \hat{H} \ket{\Psi_n} &=& E_n \ket{\Psi_n}, \qquad \hat{H} = \frac{1}{2m} \sum_{i=1}^N \pp_i^2 + \hat{V} \nn  \pp_i &=& \p_i + \frac{e}{c}{\bf A}(\r_i) , \qquad \p_i = -i \hbar \nabla_{\r_i} \label{hamiltonian} \eea where 
the potential $\hat{V}$ comprises one-body and two body terms, and is a multiplicative operator in the Schr\"odinger representation. Both the potential $\hat{V}$ and the vector potential ${\bf A}(\r)$  may depend parametrically on $\kappa$ and $\lambda$. 

I assume a singlet ground state and  irrelevant spin variables are neglected. Whenever Cartesian components need to be addressed, they are indicated with Greek subscripts; sum over repeated Cartesian indices is implicitly understood.

In the special case where $\ket{\Psi_0}$ is expressed as the Slater determinant of doubly occupied single-particle orbitals (either Hartree-Fock or Kohn-Sham), $\Omega(\kappa,\lambda)$ becomes a sum of single-particle curvatures: the expression is presented in Appendix \ref{sec:indep}.

\subsection{Microscopic current density and continuity equation} \label{sec:curr}

The most fundamental adiabatic response is the microscopic electrical current density, which explicitly displays how the electronic charge is ``dragged'' by a slow variation of the Hamiltonian.
It is by definition the expectation value of the operator \[ \hat{\bf j}(\r) = - \frac{e}{2m} \sum_{i=1}^N \{ \delta(\r-r_i)\pp_i + \pp_i \delta(\r-r_i) \} ; \label{currdef} \] the quantum-chemical literature equivalently addresses the electronic flux, whose operator differs from \equ{currdef} by the $-e$ factor.

The lowest-order adiabatic current density is, from \equ{sum2}:  \bea {\bf j}(r,t) &=& \me{\Psi_0}{\hat{{\bf j}}(\r)}{\Psi_0}  \nn &+& 2 \hbar \dot\lambda(t) \, \mbox{Im}  \sum_{n \neq 0} \frac{\ev{\Psi_0 | \hat{{\bf j}}(\r) | \Psi_n }\ev{\Psi_n | \ddb\hat{H} | \Psi_0 }}{(E_0 - E_n)^2} . \label{nafie} \eea  Expressions for the adiabatic current density---equivalent to this one---first appeared in molecular physics in the 1980s, quite independently from the Niu-Thouless result \cite{Nafie83}.

Next I am going to show that \equ{nafie} conserves continuity to leading order in the adiabaticity parameter; if the Hamiltonian varies harmonically at frequency $\omega$, continuity is conserved to first order in $\omega$. 

Given that the current in a stationary state is divergenceless, \equ{nafie} yields \[ \nabla \cdot {\bf j}(r,t) = 2 \hbar \dot\lambda(t) \, \mbox{Im}  \sum_{n \neq 0} \frac{\ev{\Psi_0 |\nabla \cdot \hat{{\bf j}}(\r) | \Psi_n }\ev{\Psi_n | \ddb\hat{H} | \Psi_0 }}{(E_0 - E_n)^2} .\label{dnafie}  \] The expression for the operator $\nabla \cdot \hat{{\bf j}}(\r)$ is derived in Appendix \ref{sec:cont}, \equ{cont1}: \[ \nabla \cdot \hat{{\bf j}}(\r)= \frac{i}{\hbar}  [ \,\hat\rho(\r) , \hat{H} \,] , \label{rapix} \] where the time-dependent Hamiltonian $\hat{H}_t$ is replaced with \equ{hamiltonian}, evaluated at $\lambda(t)$. Therein, $\hat\rho(\r)$ is the charge-density operator \[ \hat\rho(\r)= -e  \sum_{i=1}^N \delta(\r-\r_i).  \label{rho} \] By inserting \equ{rapix} in \equ{dnafie} one gets \bea \nabla \cdot {\bf j}(r,t) &=& - 2 \dot\lambda(t) \, \mbox{Re}  \sum_{n \neq 0} \frac{\ev{\Psi_0 |\hat\rho(\r) | \Psi_n }\ev{\Psi_n | \ddb\hat{H} | \Psi_0 }}{E_0 - E_n} \nn &=& - 2 \dot\lambda(t) \, \mbox{Re } \me{\Psi_0}{\hat\rho(\r)}{\partial_{\lambda}\Psi_0} , \label{dnafie2} \eea where \[ \ket{\partial_{\lambda} \Psi_0} =  \sum_{n \neq 0} \ket{\Psi_n } \frac{\ev{\Psi_n | \ddb\hat{H} | \Psi_0 }}{E_0 - E_n}  \label{parallel} \] has been exploited. Notice that the parallel-transport gauge \cite{Vanderbilt} is implicitly assumed in \equ{parallel}; the gauge choice is nonetheless irrelevant, given that \equ{dnafie2} is gauge-invariant.

The adiabatic time-derivative of the charge density obtains from the first term in \equ{sum2} only, where $\hat{O}$ is identified with the density operator $\hat\rho(\r)$. In fact the time derivative of the second term therein is proportional to $\ddot \lambda(t)$; equivalently, it is of order $\omega^2$ if the Hamiltonian varies harmonically in time. Also, for a T-invariant system, the second term in \equ{sum2}  vanishes identically (all terms in the sum are real).
One thus gets  \[ \frac{\partial \rho(\r,t)}{\partial t} = 2 \dot\lambda(t) \, \mbox{Re }\me{\Psi_0}{\hat\rho(\r)}{\partial_{\lambda}\Psi_0}  , \] which in fact coincides with minus \equ{dnafie2}.

\subsection{Molecules vs. condensed matter} \label{sec:mol}

The results presented so far apply to either an isolated molecule or to condensed matter. Once the Hamiltonian is given, the eigenvalue problem is uniquely determined when the adopted boundary conditions are specified, which amounts to a choice of the Hilbert space where the state vectors are defined.

In the molecular case one addresses the bounded ground state of an $N$-electron system. One thus requires that the wavefunction vanishes far away from the sample: such boundary conditions are dubbed ``open'' in condensed-matter theory jargon. The role of quantum geometry in some adiabatic response functions of an isolated molecule has been recently elucidated by \textcite{rap166,rap168} by adopting open boundary conditions and notations similar to the present ones; here instead I am going to address condensed-matter properties only.

One deals with solids and liquids by addressing an unbounded many-electron system within Born-von-K\`arm\`an  periodic boundary conditions (PBCs): the many-body wavefunctions are periodic with period $L$ over each electron Cartesian coordinate $r_{i\alpha}$ independently. As said above, this determines the Hilbert space where the eigensolutions of \equ{hamiltonian} are defined.

In the condensed-matter case one thus considers a system of $N$ interacting $d$-dimensional electrons in a cubic box---often called ``supercell''---of volume $L^d$, together with its periodic replicas. Both the vector potential $\A(\r)$ and the many-body potential $\hat{V}$ obey PBCs; the latter  includes the classical nuclear-nuclear repulsion; the supercell is charge-neutral. Therefore both electronic and nuclear coordinates are equivalent to angles, i.e. they are defined on a torus.

The PBCs eigenstates are normalized to one over the supercell;
the limit $N \rightarrow \infty$,  $L \rightarrow \infty$, $N/L^d = n$ constant is understood. It is not required that the system is crystalline, only that it is macroscopically homogeneous. 

\subsection{Macroscopic current density in insulators and metals} \label{sec:mcurr}

In order to address the macroscopic current it is expedient to rewrite the Hamiltonian in \equ{hamiltonian} as \[ \hat{H}_{\kk} = \frac{1}{2m} \sum_{i=1}^N \left[\p_i + \frac{e}{c}{\bf A}(\r_i) + \hbar \kk \right]^2 + \hat{V}, \label{kohn2} \] where the vector potential ${\bf A}(\r)$ summarizes all intrinsic T-breaking terms, as e.g. those due to a coupling to a background of local moments.  The ``flux'' $\kk$---cast into inverse-length dimensions for convenience---corresponds to perturbing the Hamiltonian with a vector potential  $\delta \A = \hbar c \kk /e$, constant in space. If furthermore $\kk$ is set to be constant even in time, then its occurrence is a pure gauge transformation. 

As first realized by \textcite{Kohn64}, PBCs violate gauge-invariance: in fact for a generic $\kk$ the gauge-transformed wavefunction \cite{Ballentine} obeys  Schr\"odinger equation, but no longer obeys PBCs. The genuine eigenstates of \equ{kohn2} have a nontrivial $\kk$ dependence,
and the energy may depend on $\kk$; actually, the energy is $\kk$-independent in insulators \cite{Watanabe18} and $\kk$-dependent in metals.

The macroscopic current density  ${\bf j}(t)$ is by definition the supercell average of the microscopic one, \equ{currdef}. The virtue of \equ{kohn2} is that it allows writing the macroscopic current operator as the $\kk$-derivative of the Hamiltonian: \[ \hat{\bf j} = -\frac{e}{m L^d} \sum_{i=1}^N \left[\p_i + \frac{e}{c}{\bf A}(\r_i) + \hbar \kk \right] = -\frac{e}{\hbar L^d} \partial_{\kk} \hat{H}_{\kk} . \label{current} \] This expression requires the potential $\hat{V}$ to be multiplicative, i.e. diagonal in the Schr\"odinger representation: nonlocal pseudopotentials are ruled out.

If one identifies the generic  parameter $\kappa$ in \equ{rap} with a Cartesian component of the flux $\kappa_\alpha$ in \equ{current}, then \equ{NT} immediately yields \[ j_\alpha(t) = -\frac{e}{\hbar L^d} \partial_{\kappa_\alpha} E_{\kk}  +\frac{e}{L^d} \Omega(\kappa_\alpha,\lambda) \dot\lambda(t) , \label{cm}
\] where $E_{\kk} $ is the ground-state eigenvalue of  $\hat{H}_{\kk} $; both the energy and the curvature depend parametrically on time.

The focus of this Review is on physical properties which are well defined by means of the adiabatic limit. While in insulators adiabaticity is obvious (the spectrum is gapped), the metallic case---also dealt with in the following---requires a crucial observation. Following Kohn, $\kk$-derivatives will be evaluated first, and the $L \rightarrow \infty$ limit taken afterwards. At any finite size the spectrum is gapped even in metals; the procedure warrants that the response of a metallic system to an infinitely slow perturbation is indeed adiabatic.

Equivalently, one may consider time scales. The inverse of the finite-$L$ gap sets the internal time scale of the many-electron system, while the Hamiltonian evolves in time $t$. The adiabatic limit is the $t \rightarrow \infty$ limit, and such limit is here performed {\it before} the  $L \rightarrow \infty$ one.

\section{Macroscopic polarization} \label{sec:pol}

Macroscopic polarization $\P$ is well defined for insulators only. It is the sum of an electronic and a nuclear (classical) term, similar in this to the dipole of a bounded molecule; in both cases charge neutrality is the essential requirement.

\subsection{Induced polarization} \label{sec:ind}

If the Hamiltonian is adiabatically varied by means of a parameter $\lambda(t)$, the time-derivative of $\P$ equals the macroscopic current $\j^{(\rm total)}$, whose electronic contribution is given by \equ{cm}. One therefore gets the $\lambda$-derivative of the polarization's electronic term as \[ \partial_\lambda P^{(\rm electronic)}_\alpha = \frac{j_\alpha}{\dot\lambda} = \frac{e}{L^d} \Omega(\kappa_\alpha,\lambda) . \label{pl} \]
It is worth noticing that, at the origin of polarization theory in the early 1990s (called ``modern'' at the time), a key step---due to \textcite{King93}---was indeed the identification of a polarization derivative with a Berry curvature. It was in that case a single-particle curvature, Brillouin-zone integrated; the above compact formula generalizes indeed their finding (see Appendix \ref{sec:band}).

The curvature in \equ{pl} is evaluated at $\kappa=0$ and at any $\lambda$,  while the large-$L$ limit is implicit. In some of the forthcoming developments it proves convenient to replace
\[  \left. \Omega(\kappa_\alpha,\lambda)\right|_{\kk=0} \simeq \frac{L}{2\pi}  \int_0^{\frac{2\pi}{L}} d\kappa_\alpha\; \Omega(\kappa_\alpha,\lambda) ; \label{replace} \] the reasons why are detailed next.

It has been noticed above that PBCs violate gauge invariance; therefore the state vector $\ket{\Psi_{0\kk}}$ has a nontrivial $\kk$ dependence.  Nonetheless it has been shown \cite{Watanabe18} that---for a gapped insulator---the $\kk$-dependence of the curvature is negligibly small (exponentially in $L$). One further niotices that, whenever the $\kk$ components are integer multiples of $2\pi/L$ the state vector
\[ \ket{\tilde\Psi_{\kk}} = \emi{\kk \cdot \sum_i \r_i}  \ket{\Psi_0}  \] is an eigenstate of \equ{hamiltonian} with eigenvalue $E_0$: it obeys both PBCs and Schr\"odinger equation. It coincides with the genuine $\kk$-evolved ground eigenstate 
$\ket{\Psi_{0\kk}}$ in the insulating case only; in the metallic case $\ket{\Psi_{0\kk}}$ has instead a $\kk$-dependent energy \cite{Kohn64}, and is orthogonal to $\ket{\tilde\Psi_{\kk}}$.

In view of the above the $x$-component of induced polarization may be written as  \[ \partial_\lambda P^{(\rm electronic)}_x  = \frac{e}{2\pi L^{d-1}} \int_0^{\frac{2\pi}{L}} d\kappa_x \; \Omega(\kappa_x,\lambda) . \label{pl2} \]  The end points $\kappa_x=0$ and $\kappa_x=2\pi/L$ are equivalent points, provided  the periodic gauge 
 \[ \ket{\Psi_{0 \kk_1}} = \emi{\frac{2\pi}{L} \sum_i x_i} \ket{\Psi_0}  , \quad \kk_1 = \frac{2\pi}{L}(1,0,0)  \label{periodic} \] is enforced at any $\lambda$.
 
\subsection{Polarization differences} \label{sec:diff}

Spontaneous polarization $\P$ (as e.g. in ferroelectrics) has long evaded a microscopic definition; erroneous statements still plague many textbooks. Historically, the problem was first solved in the early 1990s by addressing polarization {\it differences} instead of polarization ``itself'' \cite{rap73,King93}.

\begin{figure}[b]
\centering
\includegraphics[width=3cm]{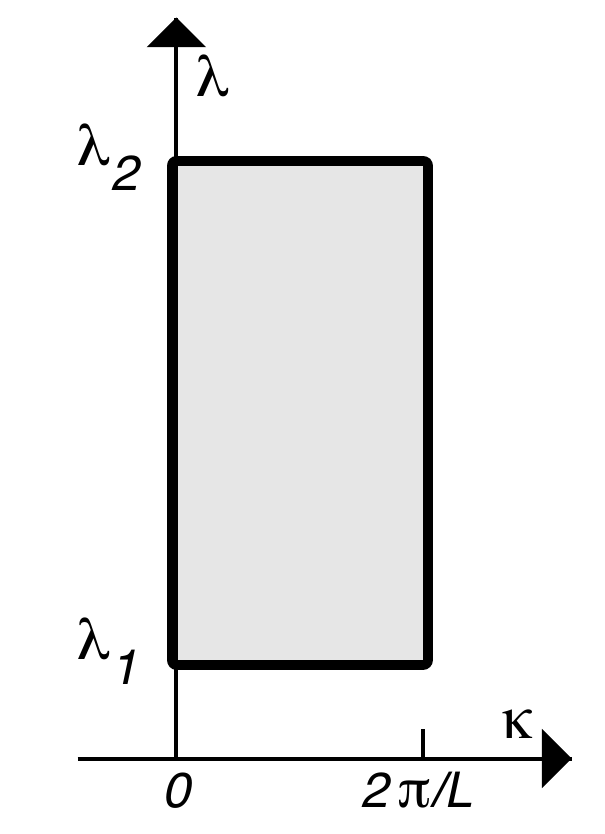} 
\caption{The integral in \equ{stokes}, evaluated on the contour by means of Stokes theorem. Because of the periodic-gauge condition the contributions from the vertical sides cancel. In the $L \rightarrow \infty$ limit each horizontal side contracts to a point, as in \equ{sp}; it thus becomes a "single-point Berry phase" \cite{rap100,rap167}.
}
\label{fig1} \end{figure}

Suppose that $\lambda_1$ and $\lambda_2$ label two different configurations of the same solid, adiabatically connected by an insulating  continuous path. Then from \equ{pl2} \bea \Delta P_x &=& P_x(\lambda_2) - P_x(\lambda_1) \label{pl3} \\ &=& \frac{e}{2\pi L^{d-1}} \int_{\lambda_1}^{\lambda_2} d \lambda \int_0^{\frac{2\pi}{L}} d\kappa_x \; \Omega(\kappa_x,\lambda) + \Delta P_x^{(\rm nuclear)} , \nonumber \eea Since the curvature, \equ{curva}, is the curl of the connection,  the integral in \equ{pl3} can be evaluated by means of Stokes theorem, as shown in Fig. \ref{fig1}.

For the sake of clarity the connection's two arguments are explicitly displayed: \[ {\cal A}_{\kappa_x}(\kappa_x,\lambda) = i \ev{\Psi_0|\partial_{\kappa_x}\Psi_0}, \quad {\cal A}_\lambda(\kappa_x,\lambda) = i \ev{\Psi_0|\partial_\lambda\Psi_0} ; \] 
\bea && \int_0^{\frac{2\pi}{L}} d\kappa_x \int_{\lambda_1}^{\lambda_2} d \lambda \; \Omega(\kappa_x,\lambda)  
\nn &=&  \int_{0}^{\frac{2\pi}{L}} d\kappa_x [\, {\cal A}_{\kappa_x}(\kappa_x,\lambda_1) - {\cal A}_{\kappa_x}(\kappa_x,\lambda_2)\,]  \nn &+&  \int_{\lambda_1}^{\lambda_2} d\lambda [\, {\cal A}_\lambda(2\pi/L,\lambda) - {\cal A}_\lambda(0,\lambda)\,]  . \label{stokes} \eea
Because of \equ{periodic}  the bottom line of \equ{stokes} vanishes; what remains is the difference of two Berry phases. We thus rewrite \equ{pl3} as \[ \Delta P_x =  - \frac{e}{2\pi L^{d-1}} [\, \gamma(\lambda_2) - \gamma(\lambda_1) \,]  +  \Delta P_x^{(\rm nuclear)} , \label{om} \] \[ \gamma(\lambda) =  \int_{0}^{\frac{2\pi}{L}} d\kappa_x \;  {\cal A}_{\kappa_x}(\kappa_x,\lambda) ; \] this many-body polarization formula owes to \textcite{Ortiz94}. Given that the Berry phase is defined modulo $2\pi$, \equ{om} is multivalued; furthermore the $1/L^{d-1}$ factor looks problematic for $d>1$. Both issues will be discussed in Sec. \ref{sec:multi}.

\subsection{Polarization ``itself''} \label{sec:itself}

Given the above, one may now switch from polarization differences to polarization ``itself'', provided one abandons the standard textbook definition of $\P$ as a vector; such change of paradigm owes to \textcite{Vanderbilt93}. Therefore $\P$ within PBCs is a lattice, i.e. a multivalued observable.

In the present many-body formulation we adopt
the expression \[ P_x =  - \frac{e}{2\pi L^{d-1}} \int_{0}^{\frac{2\pi}{L}} d\kappa_x \;  {\cal A}_{\kappa_x} + P_x^{(\rm nuclear)} . \label{multi} \] 
In the large-$L$ limit the integration segment in \equ{multi} contracts to a point: we may thus replace the Berry phase with the phase difference between the end points, as per \equ{berry}, in order to arrive at the so-called  "single-point Berry phase'' \cite{rap100,rap167}: \[ \int_{0}^{\frac{2\pi}{L}} d\kappa_x \;  {\cal A}_{\kappa_x} \simeq - \mbox{Im ln} \ev{\Psi_{0} | \Psi_{0 \kk_1} } = \mbox{Im ln} \ev{\Psi_{0} | \ei{\frac{2\pi}{L} \sum_i x_i}| \Psi_0} . \label{sp} \] The nuclear contribution can be cast as an angle as well. If $X_\ell$ is the $x$-coordinate  of nucleus $\ell$, with atomic number $Z_\ell$, the ultimate expression for the total polarization is \[  P_x =  - \frac{e}{2\pi L^{d-1}}  \mbox{Im ln} \ev{\Psi_{0} | \ei{\frac{2\pi}{L} \left(\sum_i x_i - \sum_\ell Z_\ell X_\ell \right)}| \Psi_0}  ; \label{rapix2} \] this form explicitly displays invariance by translation of the coordinate origin (owing to charge neutrality).

\subsection{Polarization as a multivalued observable} \label{sec:multi}

As said above, a basic tenet of polarization theory is that bulk polarization is a lattice, not a vector \cite{Vanderbilt93,Vanderbilt}  in the formulation of \equ{rapix2} the multivaluedness owes to the function
``Im ln''. But it is also clear that for $d>1$  \eqs{om}{rapix2} cannot be accepted as they stand in the large-sample limit: the prefactor goes in fact to zero. It has been shown by \textcite{Ortiz94} that the drawback is overcome by exploiting crystalline symmetry; an alternative proof is presented here.

For $d=3$ we consider---without loss of generality---a simple cubic lattice of constant $a$, where the supercell side $L$ is an integer multiple of $a$: $L=Ma$. When the Hamiltonian is adiabatically varied in time, the phase angle in \equ{rapix2} evolves as
 \[ \gamma(t)  =  \mbox{Im ln} \ev{\Psi_{0}(t) | \ei{\frac{2\pi}{L} \left(\sum_i x_i - \sum_\ell Z_\ell X_\ell \right)}| \Psi_0(t)}  , \] 
The current flowing across a section of area $L^2$ normal to $x$ is \[ I_x =  L^2 \dot P_x = - \frac{e}{2\pi} \dot\gamma .\] Owing to crystalline periodicity, The current $I_x$ is the sum of $M^2$ identical currents, each flowing through a microscopic section of area $a^2$; one can therefore define a reduced crystalline phase angle
$\gamma^{(\rm crystal)}$ such that $ \dot\gamma = M^2 \dot\gamma^{(\rm crystal)}$. The crystalline polarization is thus expressed in terms of $\gamma^{(\rm crystal)}$ as \[ P_x = \frac{e}{2\pi a^2} \gamma^{(\rm crystal)} , \] and is therefore defined modulo $e/a^2$.

A generic lattice is dealt with by means of a coordinate transformation \cite{rap_a12}; the bulk value of ${\bf P}$ is then ambiguous modulo $e\R/{\cal V}_{\rm cell}$, where $\R$ is a lattice vector and ${\cal V}_{\rm cell}$ is the volume of a primitive cell. 
The quantity $e\R/{\cal V}_{\rm cell}$ goes under the name of polarization ``quantum'' \cite{Vanderbilt}.

By definition, whenever a material is crystalline, a {\it uniquely defined} lattice can be associated with the real sample. The lattice is a ``mathematical construction'' \cite{Kittel},  defined---by means of an appropriate average---even in cases with correlation, finite temperature, quantum nuclei, chemical disorder (i.e. crystalline alloys, a.k.a. solid solutions)\dots, where the actual wavefunction may require a supercell (multiple of the primitive lattice cell). 
By definition a primitive cell is a minimum-volume one: this choice is mandatory in order to make ${\bf P}$ an unambiguously defined multivalued observable. 

Polarization ``itself''---also addressed as ``formal'' polarization in \textcite{Vanderbilt} reference textbook---is ill defined in noncrystalline insulators. Nonetheless liquid and amorphous materials can be polarized, in the sense that polarization {\it derivatives} (and differences) are well defined therein. Indeed the single-point Berry-phase formula, \equ{rapix2}, is routinely implemented in the framework of Car-Parrinello simulations in order to compute the infrared spectra of liquid and amorphous materials \cite{Debernardi97,Silvestrelli97,Tuckerman02}: only polarization {\it fluctuations}---and not polarization itself---are the main entry of the Kubo formula. The independent-particle version of \equ{rapix2} is given in Appendix \ref{sec:band}.

For a crystalline system of independent electrons (either Kohn-Sham or Hartree-Fock) the many-body wavefunction is the Slater determinant of Bloch orbitals. In the $M \rightarrow \infty$ limit the electronic contribution to the phase $\gamma^{(\rm crystal)}$ converges to the famous Berry-phase expression due to \textcite{King93}; this is discussed in Appendix \ref{sec:band}.

Finally we observe that the modulo ambiguity is only removed when the termination of the bounded sample is specified; it is also required that even the surfaces, as well as the bulk, are insulating \cite{Vanderbilt}. Insofar as the crystalline system is unbounded the modulo ambiguity cannot be removed.

\section{Faraday transport in electrolytes} \label{sec:Faraday}

I consider here liquid electrolytes, whose simplest example is a molten salt. The crucial requirement is that the liquid  is insulating: if the nuclei are ideally kept clamped, no electronic current flows. All of charge transport occurs then by means of an associated mass transport. In the original experiment \textcite{Faraday34} actually measured---to an astonishing precision---mass-to-charge ratios for several elements (decades before the atoms' existence became universally accepted).

In modern terms Faraday's first law of electrolysis can be recast as follows: when a macroscopic number  of nuclei of a given chemical species in the electrolytic cell passes from one electrode to the other, the transported electrical charge is an integer multiple of such number times $e$. The law is additive: it concerns each different chemical species in the cell independently. 

It is worth stressing that Faraday charges are by definition adiabatically transported charges; they {\it cannot} be identified with static charges ``belonging'' to ions of a given chemical species (contrary to a widespread incorrect belief).  Suppose we ideally take a ``snapshot'' of a liquid---or even of a solid---at a given time:  partitioning the total (nuclear $+$ electronic) charge into atomic subsets  is clearly a very ambiguous task. And in fact literally dozens of definitions for the atomic {\it static} charges have been proposed in the literature \cite{Meister94}, none of them yielding integer values. 

When a nucleus takes off from one electrode and lands on the other in it pumps some electronic charge across the cell. The electronic current is driven by the electromotive force; owing to continuity, we may address transport by monitoring the passage of nuclei and electrons across an ideal section of the cell.  

At the root of topological quantization of charge transport is a milestone \textcite{Thouless83} paper;
while this  only addresses solid-state issues, its relevance for understanding quantized charge transport in ionic liquids was pointed out shortly after in a very little quoted---and presumably also little known---paper by \textcite{Pendry84}.  Since then, a few authors have built upon their result and endorsed the topological nature of Faraday's law \cite{rap_a29,Grasselli19,rap164}.

Suppose the parameter $\lambda(t)$ entering the potential $\hat{V}$ is identified with the set of the  nuclear coordinates $\{\R_\ell(t)\}$; we start by investigating the amount of charge transported in time $T$ in a given direction across the electrolytic cell, whose circuit is closed through the battery. Here the electrons  in the liquid are described as a closed quantum system (no electrodes, no battery);  in order to allow for dc  currents it is then mandatory to adopt PBCs. Both nuclear and electronic coordinates are therefore equivalently defined on a torus. Given the Hamiltonian of \equ{kohn2} and the  periodic supercell of volume $L^3$, we address the total amount of charge transported across the supercell, say in the $x$-direction,  in time $T$. The transported nuclear charge $Q^{(\rm nuclear)}$ is obviously integer; we focus on the electronic one for the time being; from \equ{cm} one gets \bea Q^{(\rm electronic)} &=& L^2 \int_0^T dt \; j_x(t) = \frac{e}{L}  \int_0^T dt \; \Omega(\kappa_x,\lambda) \dot\lambda(t)  \nn &=& \frac{e}{L}  \int_{\lambda_1(0)}^ {\lambda_2(T) } d\lambda \; \Omega(\kappa_x,\lambda) , \eea 
Therein the curvature is evaluated at $\kk=0$, while the large-$L$ limit is implicit: we therefore exploit \equ{replace} once more to obtain
 \[  Q^{(\rm electronic)} =  \frac{e}{2\pi} \int_0^{\frac{2\pi}{L}} d\kappa_x\int_{\lambda_1(0)}^ {\lambda_2(T) } d\lambda \; \Omega(\kappa_x,\lambda) .  \label{c1}\]  Given the adopted PBCs, in general some of the nuclei have likely crossed  the $x=0$ and $x=L$ boundaries of the home supercell and have moved to some of its replicas, thus dragging some electronic charge with them; the averaged current  is then $ I = Q/T = (Q^{(\rm electronic)} + Q^{(\rm nuclear)})/T$.
 
 We are now ready to perform a gedanken experiment, first devised---at the independent-electron level---by \textcite{Pendry84}. We pick only one of the nuclei, say the $s$-th, and we drive it to the next replica cell; equivalently we drive it once round the torus. We assume that the other nuclei move out of the way in order to minimize the instantaneous energy, and that the other nuclei do not cross the supercell boundaries (except for two-way crossings). As explained above, for insulators $\kappa_x=0$ and  $\kappa_x = 2\pi/L$ are equivalent points; if furthermore
 the final configuration $\lambda_2(T)$ coincides with the initial one $\lambda_1(0)$, then the integral in \equ{c1} is over a closed surface (a torus) and 
 \[  \frac{1}{2\pi} \int_0^{\frac{2\pi}{L}} d\kappa_x\int_{\lambda_1(0)}^ {\lambda_2(T) } d\lambda \; \Omega(\kappa_x,\lambda) = C_1 ,\] where  $C_1 \in {\mathbb Z}$ is a Chern number. The Chern theorem requires the system to be insulating in the integration domain.  We thus arrive at the outstanding result \[  Q_s^{(\rm electronic)} = e \, C_1. \] The Pendry-Hodges gedanken experiment is a realization of what nowadays is generally called a ``Thouless pump'', whose general theory owes  (in the many-electron case) to \textcite{Niu84}.
 
 Some further  logical steps, first spelled out---at the independent-electron level---by \textcite{Grasselli19}, are needed in order to identify $Q_s  = e(C_1 + Z_s)$ with the Faraday charge of species $s$. 
 
 First we notice that  two nuclei  having the same atomic number accumulate the same Chern number in the period $T_{\rm c}$ of one pump cycle. In fact the Chern theorem only requires the configuration $\{\R_\ell(T_{c})\}$ to be equal to $\{\R_\ell(0)\}$.  It does not require the individual coordinates $\R_\ell(t)$ to return to themselves; exchanges of identical nuclei are allowed.
 
 A liquid configuration returning to itself is clearly a rare event;  but for a macroscopic time $T$ one  needs to consider a large number $n_{\rm c}$ of pump cycles, such that the averaged current is \[ I = \frac{e n_{\rm c}(C_1 + Z_s)}{T} . \label{cycle} \]
Consider next the general case where the nuclear trajectory $\{\R_\ell(t)\}$ ends in an arbitrary configuration  with $\{\R_\ell(T)\}$ different  from $\{\R_\ell(0)\}$. We can always imagine a further evolution such that $\{\R_\ell(T+\Delta T)\} = \{\R_\ell(0)\}$,  where $T$ is macroscopic and $\Delta T$ is microscopic;  the transported charge $\Delta Q$ is microscopic as well, and the averaged current is \[ I = \frac{e n_{\rm c} (C_1 + Z_s) +\Delta Q}{T+\Delta T} . \label{cycle2} ,\] which displays Faraday's quantization in the large-$T$ limit.

\section{Longitudinal dc conductivity} \label{sec:Drude}

The (linear) frequency-dependent conductivity tensor is by definition \[ \sigma_{\alpha\beta}(\omega) = \sigma^{(+)}_{\alpha\beta}(\omega) + \sigma^{(-)}_{\alpha\beta}(\omega)  = \frac{\partial j_\alpha(\omega)}{\partial \EE_\beta(\omega)} , \label{sigma} \] where $\EEE(\omega)$ is the macroscopic electric field; it is here partitioned into its symmetric (longitudinal) and antisymmetric (transverse) components, $\sigma^{(+)}$ and $\sigma^{(-)}$, respectively. Each of them comprises a real and an imaginary part, related to each other by the Kramers-Kronig relationship. 

Here only dc conductivity is addressed, in which case the adiabatic response theory is enough, as it will be shown in the following. It is also worth reminding that---for the reason given above (end of Sec. \ref{sec:mcurr})---the adiabatic limit is legitimate despite the fact that the spectrum of a metal is gapless in the $L \rightarrow \infty$ limit.

Next the $\lambda$ parameter in \equ{cm} is identified with $\kappa_\beta$. The second term therein is proportional to $\dot\kk$, i.e. to the electric field \[ \EEE = -\frac{1}{c}\frac{d}{dt} \delta {\bf A} = -\frac{\hbar}{e} \dot\kk \]
by means of an antisymmetric coefficient: it therefore accounts for transverse conductivity, discussed in Sec. \ref{sec:ahc}. For the time being I address the first term in \equ{cm}, indicated as $j^{(+)}_\alpha(t)$  in the following. Taking its time derivative at $\kk=0$ one gets  \[ \partial_t  j^{(+)}_\alpha(t) = -\frac{e}{\hbar L^d} \frac{\partial^2 E_0}{\partial t \partial \kappa_\alpha}  =  \frac{e^2}{\hbar^2 L^d} \frac{\partial^2 E_0}{\partial \kappa_\alpha \partial \kappa_\beta} \EE_\beta  :  \label{acce} \] this clearly shows that the many-electron system in a constant field undergoes free acceleration (in absence of extrinsic dissipation mechanisms).

In order to make contact with the existing literature \equ{acce} is rewritten as \[ \partial_t  j^{(+)}_\alpha(t) =  \frac{D_{\alpha\beta}}{\pi} \EE_\beta, \qquad D_{\alpha\beta} = \frac{\pi e^2}{\hbar^2 L^d} \frac{\partial^2 E_0}{\partial \kappa_\alpha \partial \kappa_\beta}  ,  \label{acce2} \] where $D_{\alpha\beta} $ is the Drude weight, alternatively also called adiabatic charge stiffness; clearly $D_{\alpha\beta}$ measures the inverse inertia of the many-electron system. In the special case of an electron gas (either interacting or non interacting) in a flat background potential  the Drude weight is \[ D^{(\rm free)}_{\alpha\beta} =  \frac{\pi e^2 n}{m} \delta_{\alpha\beta} , \] where $n=N/L^d$ is the density of conducting electrons. In the general case one may recast $D_{\alpha\beta}$ as  \[ D_{\alpha\beta} = \frac{\pi e^2 n^*_{\alpha\beta}}{m}, \label{n*} \] where $n^*$ acquires the meaning of the effective electron density contributing to the adiabatic current \cite{Scalapino93,rap157}.

Suppose instead that $\EEE=0$ and  all the nuclei are rigidly translated with constant velocity $\v = \dot{\bf u}$. If  $\lambda$ in \equ{cm} is identified with $u_\beta$  the total current carried by nuclei and electrons is  \[ j^{(\rm total)}_\alpha = \frac{e}{L^d} \left[ N\delta_{\alpha\beta} + \Omega(\kappa_\alpha,u_\beta) \right] v_\beta  , \label{cm2} \] since by charge neutrality the sum of the atomic numbers $Z_s$ equals the total number of electrons $N=nL^d$. 

In the reference frame of the nuclei the current is carried solely by the electrons, all moving with velocity $-{\bf v}$, and whose effective density is $n^*_{\alpha\beta}$; therefore  from \equ{n*} \[ j^{(\rm total)}_\alpha = e \,n^*_{\alpha\beta} v_\beta = \frac{m D_{\alpha\beta}}{\pi e} v_\beta  \label{zsum2} . \] We thus arrive at  the equivalent Berry-curvature form of the Drude weight: \[ D_{\alpha\beta} = D^{(\rm free)}_{\alpha\beta} +  \frac{\pi e^2}{m L^d} \Omega(\kappa_\alpha,u_\beta)  , \label{het} \] where the curvature is evaluated at $\kk=0$ and ${\bf u}=0$. A geometrical form for the Drude weight (somewhat different from the present one) was first suggested by \textcite{Hetenyi13}.
The equivalence of  \equ{het} with \equ{acce2} is double-checked  below, from \equ{dcs} onwards; it requires  T-invariance at $\kk=0$, i.e. a vanishing $\A(\r)$ in \equ{kohn2}. 

In order to Fourier-transform \equ{acce2}  one writes the longitudinal term in \equ{sigma} as \[ \sigma^{(+)}_{\alpha\beta}(\omega)  = \frac{\partial j^{(+)}_\alpha(\omega)}{\partial \EE_\beta(\omega)} =  \frac{\partial j^{(+)}_\alpha(\omega)}{\partial \kappa_\beta(\omega)}  \frac{\partial \kappa_\beta(\omega)}{\partial \EE_\beta(\omega)} , \label{sigma+} \]  and the rightmost factor is addressed first.

From $\dot\kk(t) = - e\EEE/\hbar$ one gets $\kk(t) = -e\EEE t/\hbar + \mbox{const}$. Switching to Fourier transforms $\EEE(\omega) = i\omega \hbar \kk(\omega)/e$, whose inversion yields \[ \kk(\omega) = \frac{e}{\hbar} \left[-\frac{i}{\omega} + \mbox{const} \times \delta(\omega) \right]\EEE(\omega) . \] The integration constant, as usual, is determined by imposing causality: \bea \frac{\partial \kappa_\alpha(\omega)}{\partial \EE_\beta(\omega)} &=& -\frac{e}{\hbar} \left[\frac{i}{\omega} + \pi \delta(\omega) \right] \delta_{\alpha\beta} \nn &=& -\frac{e}{\hbar}\left(\frac{i}{\omega + i\eta}\right) \delta_{\alpha\beta}  , \eea where $\eta \rightarrow 0^+$ is understood.

Replacing in \equ{sigma+} one gets  \[ \sigma^{(+)}_{\alpha\beta}(\omega) = -\frac{e}{\hbar} \frac{\partial j^{(+)}_\alpha(\omega)}{\partial \kappa_\beta(\omega)}  \frac{i}{\omega + i\eta} .\]At finite frequency, $\partial j_\alpha(\omega)/\partial \kappa_\beta(\omega)$ obtains  from time-dependent linear response theory (i.e. from a Kubo formula). Here instead one deals with an adiabatic perturbation, hence the $\kk$-derivative is taken with respect to a static $\kk$. We thus get from the first term in \equ{cm}  the dc contribution to longitudinal conductivity as: 
\[ \frac{\partial j^{(+)}_\alpha}{\partial \kappa_\beta}   =  - \frac{ e}{\hbar L^d} \frac{\partial^2 E_0}{\partial \kappa_\alpha \partial \kappa_\beta} , \] \[  \sigma^{(\rm Drude)}_{\alpha\beta}(\omega) = D_{\alpha\beta} \left[ \delta(\omega) +\frac{i}{\pi \omega} \right] . \] This expression was originally obtained by \textcite{Kohn64} by manipulating the Kubo formula. The two terms in square parenthesis obey the Kramers-Kronig relationship; the response is causal but nondissipative.

 \section{Born effective charges in insulators and metals} \label{sec:Drudem}
 
 The Born (or dynamical) effective charge tensors $Z^*_{s,\alpha\beta}$ are a staple in the theory of harmonic lattice dynamics of crystalline insulators and in the theory of charge transport in insulating liquids. 
Historically, the phenomenological definition of Born tensors is in terms of polarization induced by sublattice displacements \cite{Huang50,Cochran62}.
In the 1990s it became clear that the very concept of macroscopic polarization is based on adiabatic currents \cite{rap73,King93};
therefore the most fundamental definition of the Born tensors is based on currents as well; it has the additional virtue of applying to both insulators and metals on the same ground. Clearly, the concept of polarization does not make any sense in the latter case.

Let us suppose that $\R_s$ are the equilibrium nuclear positions; then
 the Born tensor  measures the $\alpha$-component of the macroscopic current  linearly induced by displacing the $s$-th nucleus with unit velocity in direction $\beta$, under the boundary condition that the macroscopic electric field is kept vanishing. Notice that such condition is equivalent to imposing PBCs on the potential $\hat{V}$ of the perturbed solid.
 In dimensionless form the definition is thus \[ Z^*_{s,\alpha\beta} = \frac{L^d}{e} \frac{\partial j^{(\rm total)}_\alpha}{\partial v_{s,\beta}} , \qquad {\bf v_s} = \dot\R_s , \label{zz} \] where  both electrons and nuclei contribute to ${\bf j}^{(\rm total)}$.

The geometrical expression for $Z^*_{s,\alpha\beta}$ obtains from the generic expression of \equ{cm} by identifying $\lambda$  with a nuclear coordinate $R_{s,\beta}$. The first term therein vanishes (no current flows in a static lattice); by adding the nuclear current  one gets  \[ Z^*_{s,\alpha\beta} = Z_s \delta_{\alpha\beta} + \Omega(\kappa_\alpha,R_{s,\beta}) , \label{zzz} \] where $Z_s$ is the atomic number of nucleus $s$. The curvature is evaluated at $\kk=0$ and at the equilibrium atomic positions.
The fundamental definition, \eqs{zz}{zzz}, applies as it stands to both insulators and metals, although the two cases differ in a very important respect. 

When all the nuclei are translated with the same velocity ${\bf v} = \dot{\bf u}$ the total current (due to electrons and nuclei) is  \[ j^{(\rm total)}_\alpha = \frac{e}{L^d} \left( \sum_s Z^*_{s,\alpha\beta} \right) v_\beta  \label{zsum} . \] In the insulating case no current may flow for a rigid translation of the whole solid, whence the basic relationship $\sum_s Z^*_{s,\alpha\beta} =0$, called the acoustic sum rule \cite{PCM}.

In the metallic case the electrons are left behind and some current instead flows, as per \equ{cm2}, which is indeed retrieved by summing \equ{zzz} over all the nuclei: \bea \sum_s Z^*_{s,\alpha\beta} &=&  N \delta_{\alpha\beta} + \Omega(\kappa_\alpha,u_\beta) \nn &=& N \delta_{\alpha\beta} - 2 \, \mbox{Im }  \ev{\partial_{\kappa_\alpha} \Psi_0|\partial_{u_\beta}\Psi_0}\label{zzr} ,.\eea 
From \equ{zsum2} one immediately gets the outstanding sum rule  \[ \frac{1}{L^d} \sum_s Z^*_{s,\alpha\beta}  = \frac{m}{\pi e^2} D_{\alpha\beta} = \frac{m}{\hbar^2L^d} \frac{\partial^2 E_0}{\partial \kappa_\alpha \partial \kappa_\beta}  ,\label{dcs}  \] recently found by \textcite{Dreyer22} for the special case of a crystalline system of noninteracting electrons. It is also worth noticing that the Born charges in metals are dubbed ``nonadiabatic'' therein (very much in contrast with the viewpoint adopted here). 
 
A detailed proof of \equ{dcs} goes as follows. Under the T-invariance hypothesis one has\[  \ket{\partial_{\bf u} \Psi_0} = - \sum_{i=1}^N \ket{\partial_{{\bf r} _i}\Psi_0} = -\frac{i}{\hbar}  \sum_{i=1}^N \p_i\ket{\Psi_0} = - \frac{i m}{ \hbar^2} (\partial_{\kk} \hat{H}_{\kk})\ket{\Psi_0} \nonumber  ,\] and \equ{zzr} is recast as \[   \sum_s Z^*_{s,\alpha\beta} =  N \delta_{\alpha\beta}  + \frac{2  m}{ \hbar^2} \mbox{Re }\me{\partial_{\kappa_\alpha} \Psi_0}{\partial_{\kappa_\beta} \hat{H}_{\kk}}{\Psi_0} \label{zzr2} .\] In order to compare this with the rightmost expression in \equ{dcs} one starts from the Hellmann-Feynman theorem at an arbitrary $\kk$:  \[ \partial_{\kappa_\beta} E_{\kk} = \me{\Psi_0}{{ \partial_{\kappa_\beta} \hat H}_{\kk}}{\Psi_0} . \] Taking another derivative at $\kk=0$ one gets: \[ \frac{m}{\hbar^2} \frac{\partial^2 E_0}{\partial \kappa_\alpha \partial \kappa_\beta} = N \delta_{\alpha\beta}  +  \frac{2 m}{\hbar^2} \mbox{Re }  \me{\partial_{\kappa_\alpha}\Psi_0}{{ \partial_{\kappa_\beta} \hat H}_{\kk}}{ \Psi_0} ; \] comparison to \equ{zzr2} completes the proof of \equ{dcs}.
  
In insulators the Born effective charges manifest themselves in the phonon spectra: they generate a depolarization field $\EEE$ which contributes to the restoring force at the zone center \cite{Huang50,Cochran62}. No such contribution may exist in metals in the adiabatic limit, since macroscopic fields are screened therein (Faraday-cage effect). A few calculations of Born charges in metals are available \cite{Dreyer22,Marchese24}; their possible experimental manifestations---e.g. in the case of a binary metal---are presently under discussion \cite{Hickox23}.

\section{Anomalous Hall conductivity} \label{sec:ahc}

As anticipated above, the second term in the r.h.s. of \equ{cm} yields the transverse (or Hall) conductivity at $\omega=0$, in both insulators and metals. In the latter case extrinsic effects are also relevant \cite{Nagaosa10}: they are not addressed here  (analogously they were not addressed in the longitudinal case). Both intrinsic and extrinsic Hall conductivity can be nonzero only if the system breaks T-invariance in absence of an applied magnetic field (hence the adjective ``anomalous'').

One identifies once more $\lambda$  in \equ{cm} with $\kappa_\beta$. Since for a static field $\dot\kk = - e \EEE/\hbar$, the second term in \equ{cm} immediately yields \[   \sigma^{(-)}_{\alpha\beta}(0) = - \frac{e^2}{\hbar L^d} \Omega(\kappa_\alpha,\kappa_\beta) , \label{ahc} \] where the curvature is evaluated at $\kk=0$. \equ{ahc} holds for $d=2,3$, and for both insulators and metals; it  can be alternatively derived from the appropriate Kubo formula \cite{rap165}.

In the special case of  an insulator and  $d=2$ the Hall conductivity is topologically quantized. This famous result was first established at the independent-electron level \cite{Thouless82}, and later extended to an interacting many-electron system \cite{Niu85}. The original papers did not address the anomalous case, whose theory is indeed simpler, as e.g. in \textcite{Xiao10}. 

\begin{figure}[t]
\centering
\includegraphics[width=5cm]{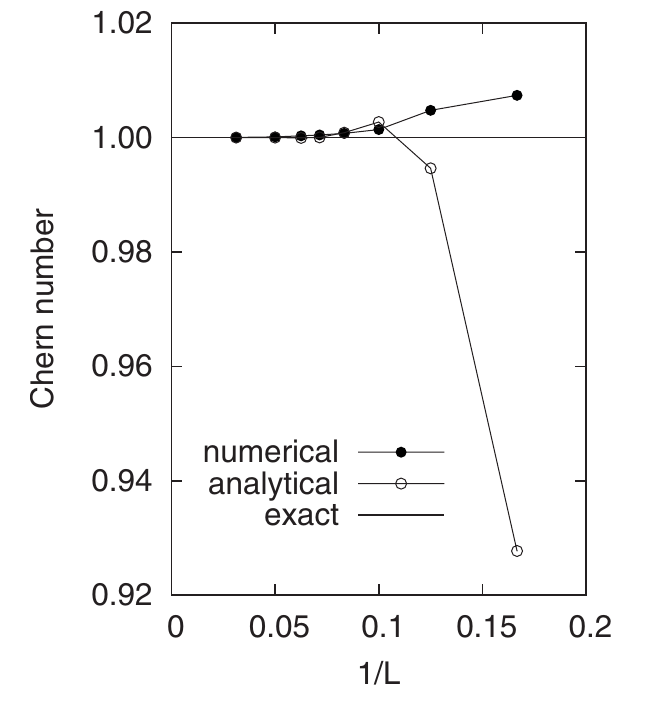} 
\caption{Convergence of \equ{TKNN} with the sample size. Simulations performed on a noninteracting \textcite{Haldane88} model Hamiltonian; the two sets of simulation points refer to two alternative discretizations of the $\kk$-derivative. Each simulation point is obtained from a single Hamiltonian diagonalization at $\kk=0$.
 After \textcite{rap135}.}
\label{fig2} \end{figure}

The form of \equ{ahc} is {\it not} quantized at finite $L$; the quantization occurs in the large-$L$ limit only. More explicitly, \equ{ahc} yields \[    \sigma^{(-)}_{xy}(0) = - \frac{2\pi e^2}{h} \lim_{L\rightarrow \infty} \frac{1}{L^2} \left. \Omega(\kappa_x,\kappa_y) \right|_{\kk=0}  = -\frac{e^2}{h} C_1, \label{TKNN} \] where $C_1 \in {\mathbb Z}$ is a Chern number; the convergence is shown in Fig. \ref{fig2} for a simple case study. 

\begin{figure}[t]
\centering
\includegraphics[width=5cm]{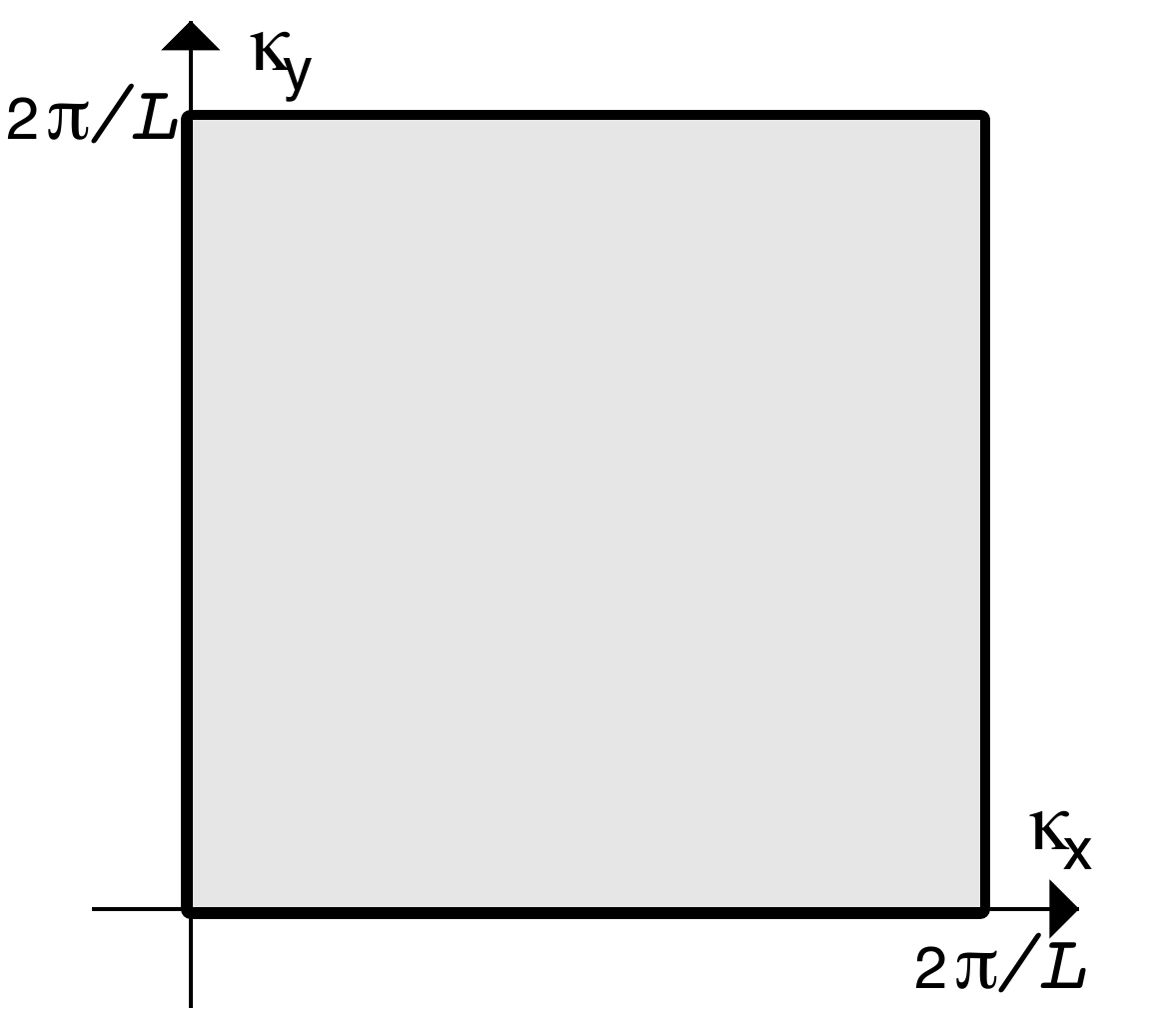} 
\caption{The integration domain in \eqs{ntw}{ntw2}; in the insulating case it is actually a torus, and the integral is quantized at any $L$. In the $L \rightarrow \infty$ limit the domain contracts to a point.
}
\label{fig3} \end{figure}

In order to prove \equ{TKNN} one writes the many-body Chern number in the formulation of \textcite{Xiao10}:
\[ C_1 = \frac{1}{2\pi} \int_0^{\frac{2\pi}{L}} d\kappa_x \int_0^{\frac{2\pi}{L}} d\kappa_y \; \Omega(\kappa_x,\kappa_y)  , \label{ntw} \] which holds at any $L$. As seen above $\kappa_\alpha=0$ and  $\kappa_\alpha = 2\pi/L$ are equivalent points in insulating systems: the integration domain of \equ{ntw}---shown in Fig. \ref{fig3}---is therefore a torus. The mean-value theorem yields \bea  \lim_{L\rightarrow \infty} \!\!\!\! \!\!\!&&\frac{1}{L^2}  \left. \Omega(\kappa_x,\kappa_y) \right|_{\kk=0} \nn &=& \lim_{L\rightarrow \infty} \frac{1}{L^2} \left(\frac{L}{2\pi}\right)^2 \int_0^{\frac{2\pi}{L}} d\kappa_x \int_0^{\frac{2\pi}{L}} d\kappa_y \; \Omega(\kappa_x,\kappa_y)  ,  \label{ntw2}\eea 
which proves \equ{TKNN}.

\section{Conclusions} \label{sec:conclu}

The observables addressed in this Review measure the adiabatic response of a condensed-matter system to a slow variation of the Hamiltonian, driven by a parameter $\lambda(t)$ entering it.  The many-body setting chosen throughout allows for compact and meaningful formulas, dealing with crystalline and noncrystalline systems on the same ground, with either interacting or noninteracting electrons. 

With the exception of macroscopic polarization, all of the observables addressed here are expressed by means of the appropriate Berry curvature; the expressions are exact for $\dot\lambda(t) \rightarrow 0$, i.e. for an infinitely slow perturbation. Higher order terms, proportional to $\ddot\lambda(t)$ and to higher derivatives are neglected within the \textcite{Niu84} approach, adopted here for all the cases discussed.

As a first step I have addressed the microscopic charge and current densities in order to fix the apparent  paradox as outlined in the Introduction. If the density $\rho(\r)$ is evaluated as the expectation value of the corresponding operator over the instantaneous ground state, its time derivative is proportional to $\dot\lambda(t)$. In order to warrant continuity the instantaneous value of the current operator is clearly not enough: one needs the current ${\bf j}(\r)$ (and its divergence) to the same order $\dot\lambda(t)$, i.e. in the adiabatic limit. This general results holds both for condensed matter and for an isolated molecule.

From such point on, this Review focusses on condensed matter;
all the addressed observables (for insulators and metals) are then defined in terms of a macroscopic electronic current. This includes observables that apparently are static, like polarization in insulators. It became clear in the 1990s that macroscopic polarization and related phenomena are indeed dynamical phenomena in the adiabatic limit \cite{Vanderbilt}. 

The theory of polarization is reviewed here in its many-body version, following the developments in \textcite{Ortiz94} and \textcite{rap100,rap162}. The polarization linearly induced by any perturbation is expressed---like most observables here---in terms of the appropriate Berry curvature. Polarization ``itself'' belongs instead to a different class: it is expressed as a Berry phase, and is a {\it multivalued} observable, defined modulo $2\pi$ in dimensionless units.

The observables addressed in the rest of this Review fall in two classes: Born and Faraday charges, and dc linear conductivities (longitudinal and transverse). 
Adiabatic response theory is adopted throughout for both insulators and metals, despite the absence of a spectral gap (in the thermodynamic limit) in the latter case. The reasons are given at the end of Sec. \ref{sec:mcurr}.

In the case  of Born and Faraday charges the parameter $\lambda(t)$ is identified with a nuclear configuration, whose motion is infinitely slow: if the nuclear motion is harmonic at frequency $\omega$, the adiabatic results are exact to order $\omega$, and do not include corrections of order $\omega^2$ and higher. One important point is worth stressing: in the present formulation the Born charges have the same formal  expression in insulators and in metals, and are adiabatic on the same ground---contrary to some opposite claims \cite{Dreyer22}. Faraday charges can be regarded as Born charges averaged over macroscopic trajectories in a liquid sample. Owing to this they are topologically quantized: their electronic contribution is a Chern number \cite{Pendry84,rap_a29,Grasselli19}; it was alternatively cast as a {\it winding number} in \textcite{rap164}.

In the case of dc conductivities---longitudinal and transverse---the $\lambda(t)$ parameter has a different meaning. The perturbation induced by a macroscopic electric field enters the Hamiltonian via a vector potential, $\r$-independent and linear in time; this is because the scalar-potential gauge violates Born-von-K\`arm\`an PBCs. Therefore  $\lambda(t)$ is identified with a Cartesian component of the vector potential (times a trivial constant); since  $\ddot\lambda(t) \equiv 0$ for a constant $\EEE$ field, adiabatic response theory is  exact to linear order in $\EEE$.
 Kubo formulas are not required to express linear conductivities---although they eventually lead to the same expressions \cite{rap165}.

One major message of  the present work is that  the longitudinal responses of a metallic solid to either a macroscopic electric field or to a rigid lattice translation measure the same observable---i.e. the inverse inertia of the many-electron system---and are expressed by the same Berry curvature.

\section*{Acknowledgments}
I have thoroughly discussed some key points of the present work with Stefano Baroni; his contribution is gratefully acknowledged.
Work supported by the Office of Naval Research (USA) Grant No. N00014-20-1-2847.

\appendix 
\section{Continuity (exact formulation)} \label{sec:cont}

In this Appendix we consider a many-body Schr\"odinger equation and one of its {\it exact} solutions: \[  \hat{H}_t \ket{\Psi} = \left(\frac{1}{2m} \sum_{i=1}^N \pp_i^2 + \hat{V}\right) \ket{\Psi}  = i \hbar \ket{\dot\Psi} , \]  where both the potential $\hat{V}$ and the vector potential ${\bf A}(\r)$  may depend on time.

In order to address the divergence of the current operator, \equ{currdef}, one needs the $\r$-gradient of $\delta(\r-\r_i)$. It is safer to evaluate how this operates on any state vector $\ket{\Psi}$ at its right: \bea && \nabla_\r \delta(\r-\r_i)\ket{\Psi} = -\{ \nabla_{\r_i} \delta(\r-\r_i) \} \ket{\Psi} =  [ \delta(\r-\r_i), \nabla_{\r_i} ] \ket{\Psi} \nn &=& \frac{i}{\hbar}[ \delta(\r-\r_i), \p_i ] \ket{\Psi} = \frac{i}{\hbar}[ \delta(\r-\r_i), \pp_i ] \ket{\Psi}. \eea One therefore gets  \bea \nabla_\r \cdot \hat{\bf j}(\r) &=& - \frac{ie}{2\hbar m} \sum_{i=1}^N \{ [\, \delta(\r-\r_i), \pp_i \,] \pp_i  + \pp_i  [\, \delta(\r-\r_i), \pp_i \,]\} \nn &=&  - \frac{ie}{2\hbar m} \sum_{i=1}^N  [\, \delta(\r-\r_i), \pp_i^2 \,], \eea 
Next we notice that for $i\neq j$ the operators $\delta(\r-\r_i)$ and $\pp_j$ commute; ergo the sum may be replaced with a double sum: \[ \nabla_\r \cdot \hat{\bf j}(\r) =  - \frac{ie}{\hbar} \sum_{i=1}^N  \left[\, \delta(\r-\r_i), \frac{1}{2m}\sum_{j=1}^N \pp_j^2 \,\right] .\] Since the potential $\hat{V}$ is multiplicative (diagonal in the Schr\"odinger representation), one eventually arrives at \[ \nabla_\r \cdot \hat{\bf j}(\r) =  - \frac{ie}{\hbar} \sum_{i=1}^N  [\, \delta(\r-\r_i), \hat{H}_t \,] = \frac{i}{\hbar}  [ \,\hat\rho(\r) , \hat{H}_t \,] ,\label{cont1} \] where $\hat\rho(\r)$ is the charge-density operator, \equ{rho}.
 
We are now equipped for addressing the time derivative of the charge density: \bea \frac{\partial \rho(r)}{\partial t} &=&\me{\Psi}{\hat\rho(\r)}{\dot\Psi} +  \me{\dot\Psi}{\hat\rho(\r)}{\Psi} \nn &=&- \frac{i}{\hbar} \me{\Psi}{  [\, \hat\rho(\r) , \hat{H}_t\, ] }{\Psi}; \label{cont2} \eea as required by continuity, \equ{cont2} coincides indeed with minus the expectation value of \equ{cont1}.
Finally I observe that the hypothesis that $\hat{ V}$ is a multiplicative operator is essential in order to warrant  continuity.

\section{Independent-electron Berry curvature} \label{sec:indep}

In the mean-field case (either Hartree-Fock or Kohn-Sham) the singlet ground state is a Slater determinant of doubly occupied single-particle orbitals $\ket{\psi_j}$ with eigenvalues $\epsilon_j$. Given that the two-body terms in $\hat{V}$ are parameter-independent, the $\kappa$-derivative entering \equ{sum} is the sum of one-body operators: \[ \dda \hat{H} = \sum_{i=1}^N O_1(\r_i) , \] and analogously for $\ddb \hat{H}$. By the Slater-Condon rules the matrix elements in \equ{sum} are converted into one-body elements of $O_1$ and $O_2$ between occupied and unoccupied orbitals; the energy differences are easily expressed as well, and \equ{sum} becomes \[ \Omega(\kappa,\lambda) = 2\,i \hspace{-0.5cm}\sum_{\stackrel{j=\mbox{occupied}}{j'=\mbox{unoccupied}}} \hspace{-0.5cm}\frac{\ev{\psi_j | O_1| \psi_{j'}  }\ev{\psi_{j'} | O_2 | \psi_j}}{(\epsilon_j - \epsilon_{j'})^2} + \mbox{cc}  , \label{sum3} \] where the factor of two accounts for spin, and ``cc' stays for complex conjugate. Straightforward manipulations yield the independent-electron  version of \equ{curva}  as \[ \Omega(\kappa,\lambda) = 2 \, \tilde\Omega(\kappa,\lambda) , \]  \[ \tilde\Omega(\kappa,\lambda) = -2\,\mbox{Im } \sum_j \theta(\mu - \epsilon_j) (\, \ev{\dda \psi_j | \ddb \psi_j} \,) ,\] where $\mu$ is the Fermi level. 

$\tilde\Omega$ can be expressed in terms of the one-body ground-state projector \[ P = \sum_j \theta(\mu - \epsilon_j) \ket{\psi_j}\bra{\psi_j} : \] \[ \tilde\Omega(\kappa,\lambda) = - i \, \mbox{Tr } \{ P\,[ \,\dda P, \ddb P \,] \} .\] This shows that $\tilde\Omega$ enjoys a generalized form of gauge invariance: it is in fact invariant by any unitary mixing of the occupied orbitals between themselves.

\section{Band-structure formulas} \label{sec:band}

For a crystalline system of noninteracting electrons (either Hartree-Fock or Kohn-Sham) the many-body wavefunction is the Slater determinant of Bloch orbitals $\ket{\psi_{j\k}} = \ei{\k \cdot \r} \ket{u_{j\k}}$ with band energies $\epsilon_{j\k}$;  the orbitals are normalized to one over the crystal cell. The discrete $\k$ vectors become a continuous variable after the $L \rightarrow \infty$ limit is taken; the observables are then expressed as Brillouin-zone (BZ) integrals in insulators, and as Fermi-volume integrals in metals.

It is easy to prove that \[ \lim_{L \rightarrow \infty} \frac{E_{\kk}}{L^d} =  \sum_j \int_{\rm BZ} \frac{d\k}{(2\pi)^d} f_j(\k) \epsilon_{j,\k\!+\!\kk} , \label{energ} \]
\bea  \lim_{L \rightarrow \infty}  &&\left.\frac{1}{L^d}\Omega(\kappa_\alpha,\lambda)\right|_{\kk=0} \nn &=& \sum_j \int_{\rm BZ} \frac{d\k}{(2\pi)^d} f_j(\k) \, \tilde\Omega_{j,\alpha\lambda}(\k) , \label{converg} \eea where  $f_j(\k)$ is the Fermi factor at $T=0$; in \equ{converg} $\tilde\Omega_{j,\alpha\lambda}(\k) $ is the Berry curvature of band $j$, defined as \cite{Vanderbilt}:
 \[ \tilde\Omega_{j,\alpha\lambda}(\k)  = -2\, \mbox{Im } \ev{\partial_{k_\alpha} u_{j\k} | \partial_\lambda u_{j\k}} . \label{omega1} \] \eqs{energ}{converg} are given per spin channel (or for ``spinless electrons''). 
 
The formulas provided so far in a many-body setting apply as they stand to the independent-electron case, by simply adopting expressions of \eqs{energ}{converg}. 
As an example, the Drude weight becomes, from \eqs{acce2}{energ}, \[ D_{\alpha\beta} = \pi e^2  \sum_j \int_{\rm BZ} \frac{d\k}{(2\pi)^d} f_j(\k) \, m^{-1}_{j,\alpha\beta}(\k) , \] where \[ m^{-1}_{j,\alpha\beta}(\k)  = \frac{1}{\hbar^2} \frac{\partial^2 \epsilon_{j\k}}{\partial k_\alpha \partial k_\beta} \] is the effective inverse mass tensor of the $j$-th band \cite{Allen06,rap157}.

Induced polarization, \equ{pl}, is transformed straightforwardly by means of \equ{converg}; the case of polarization ``itself'' requires instead a somewhat different treatment. While the Berry curvatures (either many-body or one-body) are gauge-invariant, the corresponding connections (entering polarization theory) are instead gauge-dependent.
%The electronic term in macroscopic polarization ${\bf P}$ is expressed as the BZ integral of a {\it gauge-dependent} integrand; 

Analogously to \equ{omega1}, one defines the Berry connection of band $j$:  \[ \tilde{\cal A}_{j,\alpha}(\k)  = i \ev{ u_{j\k} | \partial_{k_\alpha} u_{j\k}} . \label{conne1} \]
The conversion of the many-body formulas for $\P$ into their band-structure counterpart ---i.e. into the BZ integral of the one-body Berry connection---requires some developments, reported e.g. in \textcite{rap_a20,rap162}; only the main result is given here.
The electronic term in $ \gamma^{(\rm crystal)}$, as defined in Sec. \ref{sec:multi}, converges in the $M\ \rightarrow \infty$ limit to the  Berry phase which first appeared in the famous \textcite{King93} paper.
For a crystal with $n_{\rm b}$ doubly occupied bands the electronic term in $\P$ admits an elegant expression for a generic lattice: \[  \P^{(\rm electronic)} = -2i e \sum_{j =1}^{n_{\rm b}} \int_{\rm BZ} \frac{d\k}{(2\pi)^d} \ev{ u_{j\k} | \partial_{\k} u_{j\k}} .\] 

As said in the main text, in the noncrystalline case \equ{rapix2} is adopted as such in order to compute infrared spectra. Therein $\ket{\Psi_0}$ is---for noninteracting electrons---the Slater determinant of $N/2$ doubly occupied periodic ($\k = 0$)  orbitals $\ket{u_j}$. If one defines the ``connection matrix'' \[ S_{jj'} = \me{u_j}{\ei{\frac{2\pi x}{L}}}{u_{j'}} , \] it is easy to show  that the electronic term in \equ{rapix2} becomes \cite{rap_a20}: \[ P_x^{(\rm electronic)} = - \frac{e}{\pi L^2} \mbox{Im ln det } S . \]

%\bibliography{$HOME/inputs/huge_bib}
%\bibliography{$HOME/inputs/huge_bib,add_bib}
%\bibliographystyle{unsrt}
\bibliography{vers5}

\end{document}